# Features of the Dynamical Evolution of a Massive Disk of Trans-Neptunian Objects

V. V. Emel'yanenko

Institute of Astronomy, Russian Academy of Sciences, Moscow, Russia
e-mail: vvemel@inasan.ru

**Abstract**. Dynamical features of a massive disk of distant trans-Neptunian objects are considered in the model of the formation of small bodies in the Hill region of a giant gas-dust clump that arose as a result of gravitational instability and fragmentation of the protoplanetary disk. The dynamical evolution of orbits of small bodies under the action of gravitational perturbations from the outer planets and self-gravity of the disk has been studied for a time interval of the order of a billion years. It is shown that the secular effects of the gravitational action of a massive disk of small bodies lead to an increase in the eccentricities of the orbits of individual objects. The result of this dynamical behavior is the creation of a flux of small bodies coming close to the orbit of Neptune. The change in the number of objects surviving in the observable region of distant trans-Neptunian objects (the region of orbits with perihelion distances of $40 < q < 80$ AU and semimajor axes $150 < a < 1000$ AU), over time depends on the initial mass of the disk. For disks with masses exceeding several Earth masses, there is a tendency to a decrease in the number of distant trans-Neptunian objects surviving in the observable region after evolution for a time interval of the order of the age of the Solar system, with an increase in the initial mass. On the other hand, for most objects, orbital eccentricities decrease under the influence of the self-gravity of the disk. Therefore, the main part of the disk remains in the region of heliocentric distances exceeding 100 AU.

## INTRODUCTION

The discovery of trans-Neptunian objects (TNOs) had a huge impact on modern views on the formation of the Solar system. Although the existence of objects beyond the orbit of Neptune was predicted both in theories of planet formation (Edgeworth, 1943; Kuiper, 1951) and in studies of the origin of comets (Whipple, 1964; Fernandez, 1980), the structure of the TNO population turned out to be much more complex than previously assumed. The detection of objects moving in orbits with large eccentricities aroused great interest. The recent discovery of a family of distant TNOs moving in orbits with semimajor axes $a > 150$ AU gave new and rather unexpected information about the structure of the outer part of the Solar System. An unusual grouping of the angular elements of the orbits of these objects near certain values became the basis for the hypothesis of the existence of a distant planet producing this effect (Trujillo and Sheppard, 2014; Batygin and Brown, 2016).

Although the dynamical picture looks quite convincing (Batygin and Morbidelli, 2017), the question of the actual existence of the ninth planet of the Solar system remains open. Despite intensive searches, the planet has not yet been discovered. In the case of the existence of the ninth planet in the orbit obtained in (Batygin and Brown, 2016; Batygin et al., 2019), there are dynamical features that are difficult to match with the orbital distribution both TNOs (Shankman et al., 2017; Kavelaars et al., 2020) and comets of the Jupiter family (Nesvorny et al., 2017). It presents enormous difficulties for the explanation of the formation of such a massive and distant planet (Batygin et al., 2019).

However, without an additional disturbing body in the distant region of the Solar System, it is difficult to explain the grouping of the angular elements of distant TNOs. Therefore, in (Madigan and McCourt, 2016; Sefilian and Touma, 2019; Zderic et al., 2020), it was proposed that instead of a large planet, there is a massive disk, consisting of numerous objects of much smaller sizes. It is shown in (Emel'yanenko, 2020) that the disk of distant small bodies is a natural formation, if we assume the formation of planetesimals in the Hill regions of giant clumps arising in the early Solar system as a result of gravitational instability and fragmentation of the protoplanetary disk.

The distribution of orbits of small bodies, which was obtained in the model (Emel'yanenko, 2020), has features inherent in the distribution of the observed distant TNOs. This model examines the motion of planetesimals under the gravitational influence of two giant clumps that migrate in a disk of gas and dust and experience close encounters with each other over a relatively short period of time of several thousand years. However, in order to be observed in the modern era, the original features of the orbital distribution must persist during the lifetime of the Solar system. It is known that even for objects of the Sedna type, planetary perturbations are large enough to destroy the initial concentration of the angular orbital elements near certain values (e.g., Saillenfest et al., 2019). It is shown in (Sefilian and Touma, 2019) that a massive eccentric disk can maintain the initial distribution of the angular elements of the TNO orbits during the lifetime of the Solar system even in the presence of outer planets. However, the results of this work refer to a stationary disk with a given secular perturbing potential and do not take into account the evolution of the orbits of the objects that make up this formation.

In this work, we consider the long-term evolution of the orbits of distant TNOs, formed in accordance with the assumptions of the article (Emel'yanenko, 2020), under the influence of both perturbations from outer planets and self-gravity of the disk of objects. Since we are trying to find out what new effects are manifested in the distribution of orbits over a long period of time, in this article, in contrast to (Emel'yanenko, 2020), we study a simpler initial model, which includes consideration of the population of small bodies formed in the Hill region of a single giant clump.

MODEL OF FORMATION OF A POPULATION OF DISTANT SMALL BODIES

A system of bodies is studied, consisting of the Sun, a giant gas-dust clump, and a set of small bodies (planetesimals) located initially in the Hill region of this clump. It is assumed that a giant clump formed in the outer part of the protoplanetary disk and migrates inward as a result of interaction with this disk, according to the arguments of (Mayer et al., 2002; Vorobyov and Basu, 2005; Nayakshin, 2010; Baruteau et al., 2011; Zhu et al., 2012; Stamatellos, 2015; Vorobyov and Elbakyan, 2018). The parameters of the model correspond mainly to the work (Emel'yanenko, 2020). A giant clump with a mass equal to 17 Jupiter masses begins to move in the aphelion of the orbit with a semimajor axis $a = 110$ AU, perihelion distance $q = 100$ AU, inclination $i = 15°$, longitude of perihelion $\pi = 98°$ (orbital elements in this article are heliocentric and refer to the ecliptic plane). It is assumed that planetesimals formed in the outer part of the Hill region and are initially located in the plane of the heliocentric motion of the clump in circular orbits with a radius $r$ relative to its center. In this article, we consider planetesimals in the region $0.5R_H < r < 0.85R_H$ where $R_H = 20$ AU is the radius of the Hill sphere of the clump in aphelion. With further motion towards perihelion and migration of the clump, planetesimals begin to leave the vicinity of the clump due to the contraction of the Hill region. At the same time, their orbits experience large gravitational perturbations from the giant clump. The clump motion is considered until the moment it reaches an orbit with $q = 35$ AU. According to (Vorobyov and Elbakyan, 2018), at such a distance from the Sun, the orbit quickly rounds out due to interaction with the protoplanetary gas disk, and the clump begins to break down due to tidal effects. In any case, further the clump weakly perturbs the orbits of distant small bodies. The change in the heliocentric distance of the clump is shown in Fig. 1.

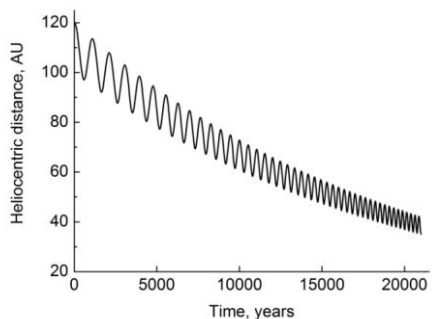

**Fig. 1.** Change in the heliocentric distance of the giant clump.

Figure 2 shows the distribution of semimajor axes and perihelion distances at the final moment of time for the orbits of small bodies with $q > 30$ AU, $150 < a < 1000$ AU and Fig. 3 shows the distribution of longitudes of perihelion and perihelion distances at the final moment of time for the same objects.

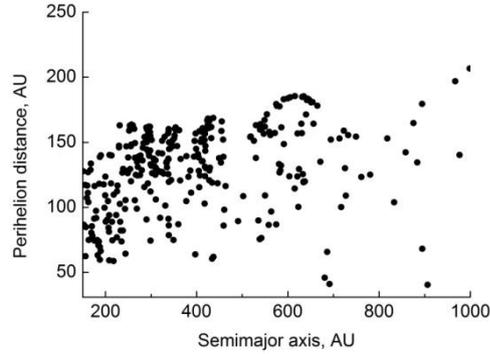

**Fig. 2.** Distribution of semimajor axes and perihelion distances for distant small bodies, arising under the action of gravitational perturbations from a giant clump.

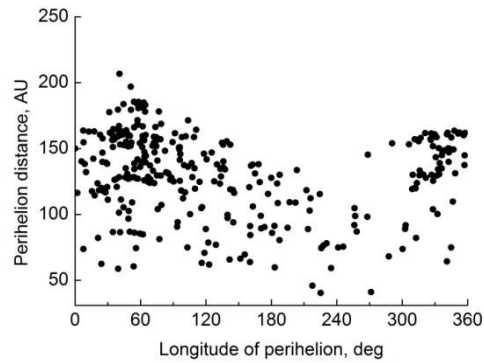

**Fig. 3.** Distribution of longitudes of perihelion and perihelion distances for distant small bodies, arising under the action of gravitational perturbations from a giant clump.

LONG-TERM EVOLUTION OF A MASSIVE DISK OF DISTANT SMALL BODIES

Objects, whose distribution of orbital elements is shown in Figs. 2 and 3, became the basis for the subsequent study of long-term evolution. From the entire set of these objects, a random sample was taken containing 340 objects. Further, it was assumed that 170 objects are massive and have the same mass, and 170 objects have zero mass. Variants were considered in which the total mass of massive objects $M_d$ has different values: 2, 6, 10, 14 and 18 $M_E$ where $M_E$ is the mass of the Earth. In addition to the self-gravity of the disk of distant objects, perturbations from the four outer planets were taken into account. Numerical integration of the equations of motion was carried out based on the use of a symplectic integrator (Emel'yanenko, 2007). In this case, the disturbing functions for Jupiter, Saturn and Uranus were averaged over the mean anomalies of the planets, and it was assumed that the orbits of these planets are circular and are in the plane of the ecliptic. This technique is often used; a more detailed description can be found, for example, in (Burns, 1976; Batygin and Brown, 2016; Zderic and Madigan, 2020). Perturbations from Neptune were fully accounted for, assuming the planet's present orbit. Integration for a given small body was terminated if $a > 1000$ AU or $q < 24$ AU as well as in the event of a collision with another object. Calculations have shown that the secular effects of the gravitational influence of the massive disk of small bodies have a predominant effect on the evolution of the orbit of an object far from the orbit from Neptune. A number of dynamical properties of the disk arising from these effects were found in (Madigan and McCourt, 2016; Madigan et al., 2018). Figure 4 shows a typical example of the evolution of the orbit of a distant object to the orbit of Neptune. The systematic decrease in the perihelion distance shows that the main changes in the orbit are not associated with close encounters of objects.

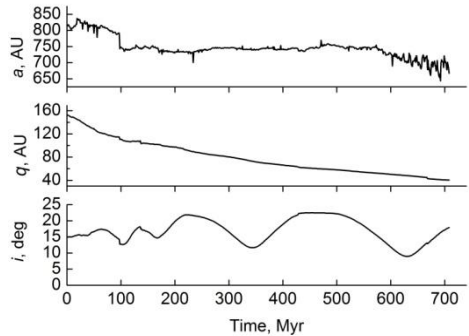

**Fig. 4.** Changes in the semimajor axis, perihelion distance and orbital inclination for a distant object penetrating to the orbit of Neptune (data with an interval of one million years are used).

The indicated dynamical behavior of individual objects leads to the creation of a flux of small bodies approaching close to the orbit of Neptune. Figure 5 shows the number of small bodies in the region $40 < q < 80$ AU, $150 < a < 1000$ AU (the observable region of distant TNOs) for different initial masses of the disk. At first, the number of objects increases, and then the flux of bodies entering this area becomes less than the flux of bodies leaving it.

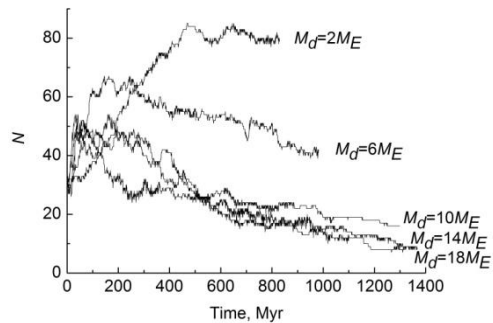

**Fig. 5.** Change in the number of small bodies $N$ in the range $40 < q < 80$ AU, $150 < a < 1000$ AU, for different initial masses of the disk (data are given with an interval of one million years). In all variants, the initial number $N = 30$.

The decrease in the number of small bodies in the observable zone of distant TNOs is associated with large perturbations from Neptune when objects approach the orbit of this planet. In this case, it is possible both the capture of objects into the inner Solar system ($q < 24$ AU) and the transition to orbits with $a > 1000$ AU. As seen in Fig. 5, the number of objects surviving in the observable region of distant TNOs strongly depends on the mass of the disk. When $M_d < 10M_E$, the rate of approach of bodies to the orbit of Neptune is very low, so most of them are preserved in the outer part of the region $40 < q < 80$ AU. When $M_d > 10M_E$ the rate of approach of small bodies to the orbit of Neptune is sufficient to remove a significant fraction of objects from this region; therefore, the number of small bodies in the observable zone of distant TNOs in this case decreases with time much faster. Computing the evolution of objects during the lifetime of the Solar system is extremely costly. Therefore, we approximated the curves in Fig. 5 in regions of decrease by exponential functions and estimated the number of small bodies after 4.5 billion years of evolution for each variant of the initial disk mass. In this approximation, the mass of objects in the region $40 < q < 80$ AU, $150 < a < 1000$ AU after 4.5 billion years is less than $0.01M_E$ for $M_d = 18M_E$, $0.04M_E$ for $M_d = 14M_E$ and $0.11M_E$ for $M_d = 10M_E$ (for smaller values of $M_d$ estimates are very uncertain).

The behavior of the distribution of longitudes of perihelion, which is the most controversial for the observed distant TNOs, is complex. Figure 6 shows the distribution of longitudes of perihelion and inclinations in the observable region of distant TNOs after 700 Myr of evolution for variants with different values of $M_d$. Although for $M_d = 10M_E$ two groupings are noticeable for longitudes of perihelion, according to the statistical Kolmogorov–Smirnov test, the probability that this distribution is not uniform is 0.5. Note that for the initial distribution of longitudes of perihelion shown in Fig. 3, this probability exceeds 0.9999. The considered dynamical process also leads to the appearance of objects in the region $30 < q < 40$ AU, $60 < a < 1000$ AU (the main part of the so-called "scattered" disk). In all versions of our model, the mass of objects in this area does not exceed $0.05M_E$ (more accurate estimates are impossible due to the insufficient number of objects in the simulation), which is consistent with the estimates obtained from observations (Gomes et al., 2008). The main part of the disk is saved in the area $q > 100$ AU, as seen in Fig. 7, which shows an example of the distribution of semimajor axes and perihelion distances for objects with $q > 30$ AU, $150 < a < 1000$ AU, after 1.2 billion years of evolution for $M_d = 10M_E$. This distribution is significantly different from the initial distribution shown in Fig. 2. The orbits of most bodies after a long evolution in a massive disk become close to circular. Such a decrease in orbital eccentricities was previously found in (Madigan and McCourt, 2016) for an eccentric ring-like disk. Our study showed that this effect also takes place for massive disks with a more complex structure.

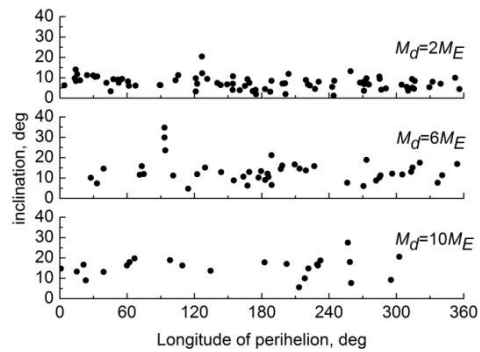

**Fig. 6.** Distribution of longitudes of perihelion and inclinations for objects with $40 < q < 80$ AU, $150 < a < 1000$ AU, after 700 million years of evolution at different values $M_d$.

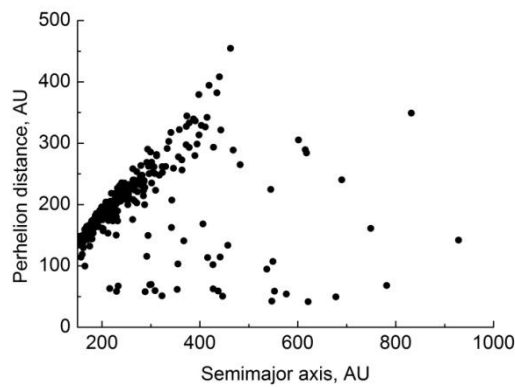

**Fig. 7.** Distribution of semimajor axes and perihelion distances for objects with $q > 30$ AU, $150 < a < 1000$ AU after 1.2 billion years of evolution at $M_d = 10M_E$.

CONCLUSIONS

The discovery of distant TNOs raised new questions about the dynamical processes that not only took place in the early Solar system, but are now taking place in the outer part of the Solar system. Modern data are still insufficient for a reliable conclusion about the structure and mass of a population of objects located far beyond the orbit of Neptune. Therefore, at present, there are various views on the origin of distant TNOs, including the hypothesis of the existence of the ninth planet. In the article (Emel'yanenko, 2020), the assumption about the origin of planetesimals in the Hill regions of migrating giant gas-dust clumps arising due to gravitational instability and fragmentation of the protoplanetary disk is considered, and it is shown that the emerging population of small bodies has a distribution of orbits similar to the observed distribution of orbits of the observed distant TNOs.

In this paper, we investigate what new features arise in the distribution of the orbits of distant TNOs, formed in accordance with the assumptions of the article (Emel'yanenko, 2020), during a time interval of the order of a billion years under the action of gravitational perturbations from the four outer planets and the self-gravity of the disk of small bodies. For this purpose, a model is considered in which small bodies originate in the Hill region of a giant clump, and the initial distribution of their orbits is created under the influence of perturbations from this clump during its migration.

It is shown that the secular effects of the gravitational action of a massive disk of small bodies lead to an increase in the eccentricities of the orbits of individual objects. This dynamical behavior leads to the creation of a flux of small bodies coming close to the orbit of Neptune. The change in the number of objects surviving in the observable region of distant TNOs ($40 < q < 80$ AU, $150 < a < 1000$ AU), over time depends on the initial mass of the disk. The estimates obtained for disks with masses exceeding several Earth masses indicate that the number of distant TNOs remained in the observable region after evolution over a time span of the order of the age of the Solar system decreases with an increase in the initial mass. On the other hand, for most objects, orbital eccentricities decrease under the influence of the self-gravity of the disk. Therefore, the main part of the disk persists in the region $q > 100$ AU.

ACKNOWLEDGMENTS

The author thanks the reviewers for useful comments. The calculations were performed using the MBC-100K supercomputer of the Joint Supercomputer Center of the Russian Academy of Sciences. The author acknowledges the support of Ministry of Science and Higher Education of the Russian Federation under the grant 075-15-2020-780 (N13.1902.21.0039).